\documentclass[aps,pra,amssymb,twocolumn,superscriptaddress,showpacs]{revtex4}

\usepackage{graphicx}
\usepackage{dcolumn}
\usepackage{bm}
\usepackage{color}
\usepackage{epstopdf}

\begin{document}
\voffset1cm

\newcommand{\beq}{\begin{equation}}
\newcommand{\eeq}{\end{equation}}
\newcommand{\barr}{\begin{eqnarray}}
\newcommand{\earr}{\end{eqnarray}}

\newcommand{\rev}[1]{{\color{red}#1}}
\newcommand{\REV}[1]{\textbf{\color{red}[[#1]]}}
\newcommand{\BLUE}[1]{\textbf{\color{blue}#1}}
\newcommand{\GREEN}[1]{\textbf{\color{green}#1}}

\newcommand{\andy}[1]{ }
\newcommand{\bmsub}[1]{\mbox{\boldmath\scriptsize $#1$}}

\def\R{\mathbb{R}}

\def\bra#1{\langle #1 |}
\def\ket#1{| #1 \rangle}
\def\sinc{\mathop{\text{sinc}}\nolimits}
\def\cV{\mathcal{V}}
\def\cH{\mathcal{H}}
\def\cT{\mathcal{T}}
\def\cM{\mathcal{M}}
\def\CW{\mathcal{W}}
\def\CR{\mathcal{R}}
\renewcommand{\Re}{\mathop{\text{Re}}\nolimits}
\newcommand{\tr}{\mathop{\text{Tr}}\nolimits}
\def\e{\mathrm{e}}
\def\ii{\mathrm{i}}
\def\d{\mathrm{d}}

\title{Generalized quantum tomographic maps }

\author{M. Asorey}
\affiliation{Departamento de F\'\i sica Te\'orica, Facultad de
Ciencias, Universidad de Zaragoza, 50009 Zaragoza, Spain}
\author{P. Facchi}
\affiliation{Dipartimento di Matematica and MECENAS, Universit\`a di Bari,
        I-70125  Bari, Italy}
\affiliation{INFN, Sezione di Bari, I-70126 Bari, Italy}
\author{V.I. Man'ko}
\affiliation{P.N. Lebedev Physical Institute, Leninskii Prospect
53, Moscow 119991, Russia}
\author{G. Marmo} \affiliation{Dipartimento di Scienze Fisiche and MECENAS,
Universit\`a di Napoli ``Federico II", I-80126  Napoli, Italy}
\affiliation{INFN, Sezione di Napoli, I-80126  Napoli, Italy}
\author{S. Pascazio} \affiliation{Dipartimento di Fisica and MECENAS,
Universit\`a di Bari,
        I-70126  Bari, Italy}
\affiliation{INFN, Sezione di Bari, I-70126 Bari, Italy}
\author{E. C. G. Sudarshan} \affiliation{Department of Physics,
University of Texas, Austin, Texas 78712, USA}

\date{\today}

\begin{abstract}
Some  non-linear  generalizations of classical Radon tomography
were recently introduced by 
M.\ Asorey \emph{et al} [Phys.\ Rev.\ A \textbf{77}, 042115 (2008)],
where the straight lines of the standard Radon map are replaced by
quadratic curves (ellipses, hyperbolas, circles) or quadratic
surfaces (ellipsoids, hyperboloids, spheres). We consider here
the quantum version of this novel non-linear approach and obtain, by
systematic use of the Weyl map, a tomographic encoding approach to
quantum states. Non-linear quantum tomograms admit a simple
formulation within the framework of the star-product quantization
scheme and the reconstruction formulae of the density operators are
explicitly given in a closed form, with an explicit construction of 
quantizers and dequantizers. The role of symmetry
groups behind the generalized tomographic maps is  analyzed in some
detail. We  also introduce new generalizations of the standard
singular dequantizers of the symplectic tomographic schemes, where the
Dirac delta-distributions of operator-valued arguments are replaced by
 smooth window functions, giving rise to the new concept of {\it  thick}
quantum tomography. Applications for quantum state measurements of
photons and matter waves are discussed.
\end{abstract}


\maketitle

\section{Introduction}
\label{sec-introd}

Quantum tomography has a long and interesting history. The first
intuition, hinting at a tomographic reconstruction of quantum
states, dates back to 1933, when Pauli asked whether it is possible
to uniquely associate quantum states with probability distributions,
as in classical statistical mechanics \cite{Pauli1,Pauli2}. Pauli's
observation was, in fact, more subtle and articulated, as he
wondered whether two given position and momentum probability
distribution functions are mathematically and physically compatible,
and whether it is possible to unambiguously reconstruct the quantum
state (the wave function) from their knowledge. The answer to
Pauli's original question is negative \cite{Reichenbach,MMSSV}, but
his general idea to associate quantum states with a set of
probability distribution functions eventually led to the basic
formulation of quantum tomography.

The tomographic representation of quantum states is based on the
Radon transform \cite{Rad1917} of the Wigner quasiprobability
distribution function \cite{Wig32}. The application to the
reconstruction of quantum states was pioneered in the 80's
\cite{Ber-Ber,Vog-Ris} and led to the first experimental
verifications in the 90's \cite{SBRF93,Mlynek96,konst}. Since then,
the tomographic approach to the analysis of quantum states has
become a booming and consolidated field of investigation
\cite{Jardabook,lwry}, leading to theoretical ideas and experimental
proposals
\cite{hradil,torino,reconstruct06,theory,theory2,pfunction,ps,bellini,allevi09},
thanks also to its important applications in quantum information
\cite{nielsenchuang,vedral}.

The original Radon transform \cite{Rad1917} maps functions defined
on a two-dimensional plane onto functions defined on a
two-dimensional cylinder. The key feature is that the transform is
invertible. There exist several important generalizations of the
Radon transform (see, e.g., \cite{Gelf,Helgason,mihlin}). More
recent analysis have focused on symplectic transforms
\cite{Mancini95}, on the deep relationship with classical systems
and classical dynamics \cite{Olga97,tomogram}, on the formalism of
star product quantization \cite{MarmoJPA} and on the study of
marginals along curves that are not straight lines
\cite{tomocurved}.

The aim of this work is to study generalizations of the Radon
transform to multidimensional phase spaces and to frameworks based
on marginals along curves or surfaces described by \emph{quadratic}
equations.  We shall study applications to both classical and
quantum systems. For classical systems we use the Radon transform of
the probability densities in phase space of a classical particle and
build the corresponding tomographic map into Radon components of the
initial probability densities. In quantum systems, we introduce a
pair of operators, the density operator and its {\it Radon transform
operator}, which are related by an invertible map. In the context of
Wigner function, these formulas provide an analogy of the Radon
transform for classical probability distributions in phase space.

The paper is organized as follows.
In Section~\ref{sec-tomogplane} we review the symplectic tomographic
approach to classical mechanics. In Section~\ref{sec-quadratic} we
provide a generalization of symplectic tomography that is based
on marginals over ellipsoids and hyperboloids and their boundary
surfaces, instead of planes and straight lines. The case of
marginals based on circles and spheres with shifted centers provides
another type of interesting generalization. We also consider, by
using a star-product framework, the generalization of Radon
transform in operator form for quantum systems in
Section~\ref{sec-star}. Center of mass tomography \cite{archi},
 symplectic tomography and its generalization to several
quantum systems is discussed in Section~\ref{sec-groupHam}, using
symplectic group Hamiltonians. A generic formulation of tomographic
approaches that make use of Hamiltonians which are linear forms in
generalized Lie algebras is briefly summarized in this section. In
Section~\ref{sec:thick} we consider {\it thick} symplectic quantum
tomography and compare it with the thick Radon transform in
Section~\ref{sec-thickRadon}. Finally, some conclusions and
perspectives are conveyed in Section~\ref{sec-concl}.

\section{Symplectic quantum tomography}
\label{sec-tomogplane}

Let $\hat{\rho}$ be a quantum state ($\hat{\rho}=\hat{\rho}^\dagger,
\hat{\rho}>0$). The quantum Radon transform (homodyne tomogram) is
given by
\begin{equation}
\CR_{\hat\rho}(X,\varphi) =\tr \hat{\rho}\ \delta\left(X \hat{\mathbb{I}}
-\hat{q} \cos\varphi - \hat{p} \sin \varphi \right) ,
 \label{homodyne}
\end{equation}
with $\varphi\in [0,2\pi]$.
In terms of the Wigner function
\begin{equation}
W(p,q)=\int \langle q-\frac{\xi}{2} | \hat{\rho} | q +\frac{\xi}{2} \rangle\; \d\xi,
\quad  \int W(p,q)   \; \d p\, \d q = 2\pi ,
\label{eq:wigner}
\end{equation}
the above expression reads
\barr
\label{eq:radon}
\CR_{\hat\rho}(X,\varphi )&=& \frac{1}{2\pi}\int \d p\,\d q\; W(p,q)
\nonumber\\ &&\times
\delta \left( X-q \cos \varphi - p \sin \varphi \right).
\earr
The quantum symplectic \cite{MarmoPL} (or $M^2$ \cite{FLP}) transform
is a generalization of the quantum Radon transform (\ref{homodyne}) and reads
\begin{eqnarray}
\CW_{\hat{\rho}}(X,\mu,\nu) &=& \left\langle \delta(X\hat{\mathbb{I}}-\mu \hat{q}-\nu
\hat{p})\right\rangle_\rho \nonumber\\
& =& {\rm Tr}\, \hat{\rho}\  \delta\left(X \hat{\mathbb{I}}-\mu \hat{q}-\nu \hat{p}\right),
\label{eq:radondef}
\end{eqnarray}
where $\mu$ and $\nu$ are real parameters. The information content
of the two formulations is identical and is expressed by the
relation
\begin{equation}
\CW_{\hat\rho}(X, r \cos\varphi , r \sin \varphi) =
\frac{1}{r}\, \CR_{\hat\rho}\left(\frac{X}{r},\varphi\right),
\label{eq:m2radon}
\end{equation}
valid for any $r>0$. Equation~(\ref{eq:m2radon}) is an easy
consequence of the fact that the Dirac distribution is positive
homogeneous of degree $-1$. The two formulations (\ref{homodyne})
and (\ref{eq:radondef}) may differ in practice (the latter being
easier to invert \cite{FLP}) and cease to be equivalent when the Dirac
delta-function is replaced by a finite window function
\cite{tomothick}. In the present article, we shall focus on the
symplectic version (\ref{eq:radondef}).

Equation (\ref{eq:radondef}) can be rewritten as the ``classical"
tomogram of the Wigner function
\begin{equation}
\CW_{\hat{\rho}}(X,\mu,\nu) = \frac1{2\pi}\int W(p,q) \,
\delta\left(X -\mu {q}-\nu {p}\right) \d p\, \d q,
\label{eq:radondos}
\end{equation}
whose inverse transform reads
\begin{eqnarray}
W{({p},{q})}= \frac1{2\pi}  \int \CW_{\hat{\rho}}(X,{\mu},{\nu}) \,
{\rm e}^{\ii (X-{\mu}{q}-{\nu}{p})}\; \d X\,\d{\mu}\, \d{\nu},
\label{invradon}
\end{eqnarray}
or, in terms of the density matrix,
\begin{eqnarray}
\hat{\rho}= \frac1{2\pi}  \int \CW_{\hat{\rho}}(X,{\mu},{\nu})
{\rm e}^{\ii (X \hat{\mathbb{I}}-{\mu}\hat{q}-{\nu}\hat{p})}\; \d X\,\d{\mu}\, \d{\nu}.
\label{invradon2}
\end{eqnarray}
For completeness and for future convenience, we notice that the
inversion formula of the Radon transform (\ref{homodyne}) is easily
obtained from Eqs.\ (\ref{eq:m2radon}) and (\ref{invradon2}) and
reads
\begin{eqnarray}
\hat\rho  &=&
\frac{1}{2\pi }
\int \d X \int_0^\infty   \d r \int_0^{2\pi} \d \varphi\; \CR_{\hat\rho} (X,{\varphi })
\nonumber \\
&&\times r\, \e^{\ii r(X \hat{\mathbb{I}} -\hat{q} \cos\varphi - \hat{p} \sin \varphi )}.
 \label{im2sim}
\end{eqnarray}

The generalization of (\ref{eq:radondos}) to the multimode case can be achieved in several
ways. Let us consider the case of two modes for simplicity. One can either introduce the ordinary
tomogram
\begin{eqnarray}
& &\CW_{\hat{\rho}}(\vec{X},\vec{\mu},\vec{\nu})
\nonumber\\
& & = \left\langle
 \delta(X_1 \hat{\mathbb{I}}-\mu_1 \hat{q}_1-\nu_1 \hat{p}_1)\,
 \delta(X_2 \hat{\mathbb{I}}-\mu_2 \hat{q}_2-\nu_2 \hat{p}_2)
\right\rangle , \nonumber\\
\label{eq:qradonmulti}
\end{eqnarray}
where $\vec{X}=(X_1,X_2)$, $\vec{\mu}=(\mu_1,\mu_2)$ and
$\vec{\nu}=(\nu_1,\nu_2)$, or, alternatively, the ``center-of-mass"
tomogram
\begin{equation}
\CW_{\rm cm}(X,\vec{\mu},\vec{\nu})\! =\! \left\langle
 \delta(X \hat{\mathbb{I}}-\mu_1 \hat{q}_1-\nu_1 \hat{p}_1-\mu_2 \hat{q}_2-\nu_2 \hat{p}_2)
\right\rangle.
\label{eq:qradoncom}
\end{equation}
The Weyl map (\ref{eq:wigner}) provides again the possibility  of expressing
(\ref{eq:qradonmulti}) and  (\ref{eq:qradoncom}) as ``classical"
tomograms
\begin{eqnarray}
\nonumber
 \CW_{\hat{\rho}}(\vec{X},\vec{\mu},\vec{\nu})\! & &=
\frac{1}{(2\pi)^2} \int \d\vec{p}\, \d\vec{q} \; W(\vec{p},\vec{q})  \\
&&\delta\left(X_1 -\mu_1 {q_1}-\nu_1 {p_1}\right)
 \delta\left(X_2 -\mu_2 {q_2}-\nu_2 {p_2}\right) ,
\nonumber\\
\label{eq:radonthree}
\end{eqnarray}
\begin{eqnarray}
 \CW_{\rm cm}({X},\vec{\mu},\vec{\nu})&=&\frac{1}{(2\pi)^2} \int \d\vec{p}\, \d\vec{q} \; W(\vec{p},\vec{q})
 \nonumber
\\
&& \delta\left(X -\mu_1 {q_1}-\nu_1 {p_1}-\mu_2 {q_2}-\nu_2 {p_2}\right),
\nonumber\\
\label{eq:cradoncom}
\end{eqnarray}
where $W(\vec{p},\vec{q})$ is the Wigner function of the two-mode
system, normalized to $(2\pi)^2$.

The inverse transform reads
\begin{equation}
W(\vec{p},\vec{q}) =  \int \CW_{\rm cm}(X,\vec{\mu},\vec{\nu}) \,
{\rm e}^{\ii (X-\vec{\mu}\cdot\vec{q}-\vec{\nu}\cdot\vec{p})}\frac{\d X\,\d\vec{\mu}\, \d\vec{\nu}}{(2\pi)^2},
\label{eq:iradoncom}
\end{equation}
for the center-of-mass tomogram, and
\begin{equation}
W(\vec{p},\vec{q})=  \int \CW_{\hat{\rho}}(\vec{X},\vec{\mu},\vec{\nu}) \,
{\rm e}^{\ii (X_1+X_2-\vec{\mu}\cdot\vec{q}-\vec{\nu}\cdot\vec{p})}\frac{\d\vec{X}\,\d\vec{\mu}\, \d\vec{\nu}}{(2\pi)^2},
\label{eq:iradonsim}
\end{equation}
for the ordinary symplectic tomogram. The inversion formulae for the
density matrix are
\begin{equation}
\hat{\rho}=  \int \CW_{\hat{\rho}}(\vec{X},\vec{\mu},\vec{\nu}) \,
{\rm e}^{\ii (X_1\hat{\mathbb{I}} +X_2 \hat{\mathbb{I}} -\vec{\mu}
\cdot\hat{\vec{q}}-\vec{\nu}\cdot\hat{\vec{p}})} \frac{\d\vec{X}\,\d\vec{\mu}\, \d\vec{\nu}}{(2\pi)^2},
\label{eq:rhoradonsim}
\end{equation}
for the symplectic case, and
\begin{equation}
\hat{\rho}=  \int \CW_{\mathrm{cm}}(X,\vec{\mu},\vec{\nu}) \,
{\rm e}^{\ii (X \hat{\mathbb{I}} -\vec{\mu}\cdot\hat{\vec{q}}-
\vec{\nu}\cdot\hat{\vec{p}})}\frac{\d{X}\,\d\vec{\mu}\, \d\vec{\nu}}{(2\pi)^2}
\label{eq:rhoradonsim2}
\end{equation}
for the center-of-mass tomography. The advantage of the symplectic
tomogram is that it permits the splitting into subsystems for
separable states, which is not possible in the center-of-mass
tomogram. This property might be very relevant for the fast
reconstruction of non-entangled states.

In the case of $M$ subsystems with their own centers of mass, 
tomography is defined by
\begin{eqnarray}
\CW_{\rm cm}(X_1,&&\!\!\!\!\!\!\vec{\mu}_1,\vec{\nu}_1, X_2,\vec{\mu}_2,
\vec{\nu}_2,\dots, X_M,\vec{\mu}_M,\vec{\nu}_M) \nonumber\\
&&=\left\langle \prod_{i=1}^{M} \delta\left(X_i \hat{\mathbb{I}}-\vec{\mu}_i \cdot
\hat{\vec{q}}_i-\vec{\nu}_i \cdot \hat{\vec{p}}_i\right)
\right\rangle,
\label{eq:qnradoncom}
\end{eqnarray}
where  $\vec{\mu}_i=(\mu_{i 1},\mu_{i 2},\dots,\mu_{i N_i})$,
$\vec{\nu}_i=(\nu_{i 1},\nu_{i 2},\dots,\nu_{i N_i})$, with
$i=1,2,\dots, M$ and
\begin{equation}
\sum_{i=1}^M N_i=N.
\label{eq:sum}
\end{equation}
The inversion formula is given by
\begin{eqnarray}
\hat{\rho}&=& \int \CW_{\rm cm}(X_1,\vec{\mu_1},\vec{\nu_1},
X_2,\vec{\mu}_2,
\vec{\nu}_2,\dots, X_M,\vec{\mu}_M,\vec{\nu}_M)\nonumber\\
&&{\rm e}^{\ii \sum_j(X_j \hat{\mathbb{I}}-\vec{\mu}_j \cdot \hat{\vec{q}}_j-\vec{\nu}_j \cdot \hat{\vec{p}}_j)}
\frac{\d\vec{X}\,\d\vec{\mu}\, \d\vec{\nu}}{(2\pi)^M}.\nonumber
\label{eq:rhonradonsim}
\end{eqnarray}
The symplectic tomogram can also be constructed in the case of multimodal systems.

\section{Quadratic tomograms}
\label{sec-quadratic}

One of the main objectives of this article is to study generalizations of the quantum Radon
transform to marginals along curves or surfaces described by quadratic, rather than linear
equations.
The solution of the classical problem was obtained in Ref.~\cite{tomocurved}.
Given an $M \times M$ symmetric operator $B$ and two $M$-dimensional vectors $\vec x$ and  $\vec \alpha$, let
\begin{equation}
\label{eq:quartic}
X=(\vec x- \vec \alpha) \cdot B (\vec x-\vec \alpha) ,
\end{equation}
where $\vec x\cdot \vec y$ denotes the scalar product of vectors
$\vec x$ and $\vec y$.
The classical tomogram of a function $f (\vec x)$ reads
\begin{eqnarray}
 \omega_f(X,\vec \alpha) = \int_{\mathbb{R}^m}  f(\vec x)\,
 \delta \left(X-(\vec x- \vec \alpha)\cdot B (\vec x- \vec \alpha)\right) {\d \vec x}.
 \nonumber\\
\label{new}
\end{eqnarray}
Observe that the $\delta$-function is supported on the quadrics
(ellipsoids, hyperboloids, etc., that can also have degeneracies)
defined by Eq.\ (\ref{eq:quartic}). The inverse map is
\begin{eqnarray}
 f(\vec x)  &=&\displaystyle\frac{|\det B|}{\pi^m } \int_{\mathbb{R}^{m+1}}
 \d X\, \d \vec \alpha  \, \omega_f(X,\vec \alpha)
 \nonumber\\
 && \qquad \times\displaystyle{\rm e}^{\ii\left(X-(\vec x- \vec \alpha) \cdot B (\vec x- \vec \alpha)\right)}.
\label{eq:inv}
\end{eqnarray}

The solution of the quantum problem is similar. With a slight abuse
of terminology, we shall speak of multidimensional phase spaces and
quadratic ``Hamiltonians" of the type
\begin{equation}
\hat{H}=\frac12 \hat{\vec{Q}} \cdot B \hat{\vec{Q}} + \vec{C}\cdot \hat{\vec{Q}},
\label{quad}
\end{equation}
where
\begin{equation}
\hat{Q}_j=\hat{p}_j, \qquad  \hat{Q}_{N+j}=\hat{q}_j, \qquad
(j=1,\dots,N),
\end{equation}
are the  momentum and position operators, and $\vec{C}$ and $B$
denote an $2N$-dimensional vector and  a $2N\times 2N$ symmetric
matrix, which parameterize the different types of the quadratic
Hamiltonian. Observe that since we are working in a symplectic
framework, $M \to 2N$.

The quantum counterpart of the classical tomogram (\ref{new}) reads
\begin{equation}
\CW(X,\vec{\mu},\vec{\nu})=\left\langle \delta\left(X \hat{\mathbb{I}}
- \hat{\mathbb{H}}_{\vec{\mu} \vec{\nu}}\right)\right \rangle,
\end{equation}
where
\begin{equation}
 \hat{\mathbb{H}}_{\vec{\mu} \vec{\nu}}= \frac12 (\hat{\vec{Q}}-
\vec{r})\cdot B\, (\hat{\vec{Q}}-\vec{r}) + \vec{C}\cdot (\hat{\vec{Q}}-\vec{r})
\label{eqcurve}
\end{equation}
and $\vec{r}=(\vec{\mu}, \vec{\nu})$ is a $2N$-dimensional vector.
The coordinate $X$ plays the role of ``energy". The Hamiltonian can
be degenerate, depending on the type of eigenvalues of the $2N\times
2N$ symmetric matrix $B$ \cite{tomothick}. The parameters
$\vec{\mu}$ and $ \vec{\nu}$ have the meaning of shift parameters of
the centers of the operatorial quadratic curves (surfaces).

From the Hamiltonian (\ref{eqcurve}) one can derive a quantum
tomographic map similar to that obtained in the classical case, by
using the Wigner function:
\begin{equation}
\CW(X,\vec{\mu},\vec{\nu})=\int\, \frac{\d\vec{p}\, \d\vec{q}}
{(2\pi)^N}\  \delta\left(X  - \mathbb{H}_{\vec{\mu} \vec{\nu}}\right)W(\vec{q},\vec{p}),
\label{wig}
\end{equation}
where the Hamiltonian $\mathbb{H}_{\vec{\mu} \vec{\nu}}$ is given by
(\ref{eqcurve}) with the operators $\hat{\vec{q}}$ and
$\hat{\vec{p}}$ replaced by c-numbers ${\vec{q}}$ and ${\vec{p}}$.
Since Eq.\ (\ref{wig}) is equivalent to Eq.\ (\ref{new}), the
reconstruction function follows directly from Eq.\ (\ref{eq:inv})
with the replacement $B\to \frac12 B$, namely,
\begin{eqnarray}
\hat{\rho} &=&  \int \CW(X,\vec{\mu},\vec{\nu})\,  \e^{\ii(X\hat{\mathbb{I}}-
\hat{\mathbb{H}}_{\vec{\mu} \vec{\nu}})} \left|\det {B}{}\right|
\frac{\d X \d\vec{\mu}\, \d\vec{\nu}}{(2\pi)^N} . \quad \label{eq:invgen}
\end{eqnarray}

The generalization to multipartite systems is straightforward. If we
have $M$ subsystems we can define the multipartite tomogram
\begin{eqnarray}
& & \CW_{\mathrm{cm}}({\vec
X},\vec{\mu},\vec{\nu})\!\!=\!\!\int\!\!W(\vec{q},\vec{p}) \prod_{j=1}^M\,    \delta\left(X_j
\hat{\mathbb{I}}
- \hat{\mathbb{H}}_{\vec{\mu}_j \vec{\nu}_j}\right) \frac{\d\vec{p}_j\,
\d\vec{q}_j}
{(2\pi)^{N_j}},\nonumber\\
& & \vec
X=(X_1,X_2,\ldots, X_M),\quad
\vec\mu=(\vec{\mu}_j),\quad\vec\nu=(\vec{\nu}_j),
\label{wig1}
\end{eqnarray}
whose inverse transform is
\begin{eqnarray}
\hat{\rho} &=&  \int \CW_{\mathrm{cm}}(\vec X,\vec{\mu},\vec{\nu})\,
\prod_{j=1}^M\, \e^{\ii(X_j\hat{\mathbb{I}}-
\hat{\mathbb{H}}{\vec{\mu}_j \vec{\nu}}_j)} \nonumber\\
&&\times \left|\det {B_j}\right|
\frac{\d X_j\, \d\vec{\mu}_j\, \d\vec{\nu}_j}{(2\pi)^{N_j}} . \quad
 \label{eq:invgenmulti}
\end{eqnarray}

\section{Star-product framework}
\label{sec-star}

It is interesting to see how the transform with quadratic curves in
phase space can be written in the form of a star-product quantization
\cite{MarmoJPA}.  Any star-product scheme is aimed at constructing a
bijective map of operators $\hat{A}$ acting on a Hilbert space $\cH$
and the space of functions  $f_A$ defined on a manifold $\cM$. The
map can be constructed in terms of two operator-valued functions
$\hat{D}$ and  $\hat{U}$  defined on $\cM$ and called ``quantizers"
and ``dequantizers", respectively. These two functions satisfy the
identity
\begin{equation}
{\rm Tr} \left(\hat{D}(x) \hat{U}(y)\right) =\delta(x-y),
\end{equation}
for every $x, y \in \cM$. The map is then defined by the formula
\begin{equation}
f_A(x)= {\rm Tr} \left(\hat{A}\, \hat{U}(x)\right),
\end{equation}
and the inverse map is defined by
\begin{equation}
\hat{A}= \int  f_A(x)  \hat{D}(x)\, \d x .
\end{equation}
The star product is defined on the space of functions $f_A$ by
\begin{equation}
f_A \star f_B=f_{A B},
\end{equation}
The kernel of this associative star product is then given by
\begin{equation}
K(x_1, x_2, x_3)= {\rm Tr} \left( \hat{D}(x_1)  \hat{D}(x_2)  \hat{U}(x_3)\right)
\end{equation}
and satisfies
\begin{equation}
f_A \star f_B(x)= \int f_A(x_1)\, f_A(x_2)\, K(x_1,x_2,x) \; \d x_1 \d x_2 .
\end{equation}

The key observation is that the tomogram can be interpreted in terms of a dequantizer
\begin{equation}
\hat{U}(X,\vec{\mu},\vec{\nu})= \delta\left(X \hat{\mathbb{I}} - \hat{\mathbb{H}}_{\vec{\mu} \vec{\nu}}\right),
\label{dequantizer}
\end{equation}
and a quantizer
\begin{equation}
\hat{D}(X,\vec{\mu},\vec{\nu})= \frac{\left|\det B\right|} {(2\pi)^N} \,
 \e^{\ii(X\hat{\mathbb{I}}-\hat{\mathbb{H}}_{\vec{\mu} \vec{\nu}})}.
\label{quantizer}
\end{equation}
Thus, for an arbitrary observable $\hat{A}$, one can introduce the
quadratic tomogram
\begin{equation}
\CW_A(X,\vec{\mu},\vec{\nu})= {\rm Tr }\hat{A}\, \delta\left(X \hat{\mathbb{I}}
- \hat{\mathbb{H}}_{\vec{\mu} \vec{\nu}}\right),
\label{observ}
\end{equation}
Also one can reconstruct the observable operator from its quadratic tomogram
\begin{eqnarray}
\hat{A} &=&  \int \CW_A(X,\vec{\mu},\vec{\nu})\, \hat{D}(X,\vec{\mu},\vec{\nu})\;
\d X \, \d\vec{\mu}\,  \d\vec{\nu} . \quad \label{eq:invobs}
\end{eqnarray}
There exist also tomograms  based on shifts of quadratic curves in
phase space, but, as in Eq.\ (\ref{eq:radondef}) for symplectic
tomograms, we used instead rescaled position and momentum operators.

One can combine both types of tomograms, derived by shifts and
rescaling of variables, obtaining a new class of tomograms, which
are the quantum version of the classical tomograms defined by
\cite{tomocurved}
\begin{eqnarray}
\omega_f(X,\vec \mu, \vec\nu)\! =\!\! \int_{\mathbb{R}^{2n}}
\delta \left(X-\vec \mu\cdot \vec q- \nu (\vec q, \vec p) \right)
 f(\vec q, \vec p) {d \vec q\ d \vec p },
\nonumber\\
\label{eq:exf2}
\end{eqnarray}
where $\vec p$ and $\vec q$ are vectors in $\R^n$ and
\begin{equation}
\nu (\vec q, \vec p)=\sum_{j=1}^n \nu_j q_j  p_j.
\end{equation}
This map corresponds to a deformation of the standard
multidimensional Radon transform by means of the following
diffeomorphism of $\mathbb{R}^{2n}\backslash\bigcup_{j}\{(\vec q,
\vec p): q_j=0\}$ \beq (q_i,p_j)\mapsto (x_i,y_j)=\left({q_i}, {q_j}
p_j\right), \eeq whose Jacobian is \beq J(\vec q, \vec
p)=\left|\frac{\partial(\vec x, \vec y)}{\partial(\vec q, \vec
p)}\right|=\prod_{j=1}^n| q_j | . \eeq The inverse map is given by
\begin{eqnarray}
 f(\vec q,\vec p)  &=&\int_{\mathbb{R}^{2n+1}} \frac{dX\, d \vec \mu\,
 d \vec \nu}{(2\pi)^{2n}}\,\omega_f(X,\vec \mu, \vec \nu)
\nonumber\\
&& \quad \times \prod_{j=1}^n| q_j |\,  {\rm e}^{i\left(X-\vec
\mu\cdot \vec q-\nu (\vec q,\vec p) \right)} . \label{eq:inv2}
\end{eqnarray}
This corresponds to the higher-dimensional generalization of the
Bertrand--Bertrand tomography \cite{Ber-Ber}.

The quantum extension is straightforward. One introduces the
Hamiltonian
\begin{eqnarray}
\hat{\mathbb{H}}_{\xi\nu}(\hat{\vec{q}},\hat{\vec{p}}) &=&
\sum_{i=1}^N \xi_i\hat{q}_i +  \frac{1}{2}\sum_{j=1}^N
\nu_j(\hat{q}_j \hat{p}_j +\hat{p}_j\hat{q}_j). \label{eq:newham}
\end{eqnarray}
The dequantizer operator
\begin{equation}
\hat{U}(X,\vec{\xi},\vec{\nu})= \delta\left(X \hat{\mathbb{I}} -
 \hat{\mathbb{H}}_{\vec{\xi} \vec{\nu}}(\hat{\vec{q}},\hat{\vec{p}})\right)
\label{dequantizerbis}
\end{equation}
yields the quantum tomogram
\begin{equation}
{\CW}(X,\vec{\xi},\vec{\nu})= {\rm Tr }\,\hat{\rho}\, \delta\left(X \hat{\mathbb{I}}
- \hat{\mathbb{H}}_{\vec{\xi} \vec{\nu}}(\hat{\vec{q}},\hat{\vec{p}})\right),
\label{observ2}
\end{equation}
and corresponds to the  quantizer
\begin{equation}
\hat{D}(X,\vec{\xi},\vec{\nu})= \prod_{j=1}^N| \hat{q}_j | \,
\e^{\ii (X\hat{\mathbb{I}}-\hat{\mathbb{H}}_{\vec{\xi}
\vec{\nu}}(\hat{\vec{q}},\hat{\vec{p}}))}. \label{quantizerbis}
\end{equation}
This permits to recover the quantum state from  its tomogram.
The modulus of the operator is defined by the Weyl quantization of the symbol $|q_j|$.

\section{Group Hamiltonians}
\label{sec-groupHam}

The applications considered so far involve position and momentum
coordinates. In the linear and quadratic case, the Hamiltonians from
which we constructed the quantizers and dequantizers can be written
in the form
\begin{eqnarray}
\hat{H}_{g_\alpha}= \sum_{\alpha} (g_o)_\alpha\hat{L}_\alpha,  \label{eq:renewham}
\end{eqnarray}
where $\hat{L}_\alpha$ are generators of Lie groups. In the examples
considered in this article, they belong to representations of the
inhomogeneous Lie group ISp$(2N,\mathbb{R})$ and its subgroups. 
The group ISp$(2N,\mathbb{R})$ is defined as the group of affine transformations 
of $\mathbb{R}^{2N}\equiv T^\ast \mathbb{R}^{N}$
which preserve its natural symplectic form. It the semi-direct product of the 
translation group and the
linear symplectic group.
The reason for this choice is
because this group acts on the Heisenberg-Weyl group as a group of automorphisms. 
Within this
framework, the delta function is expressed as
\begin{equation}
\delta\left(X\hat{\mathbb{I}} - \sum_{\alpha}
(g_o)_\alpha\hat{L}_\alpha\right)=\frac1{2\pi}\int \d t \, {\rm
e}^{\ii t (X \hat{\mathbb{I}}- \sum_{\alpha}
(g_o)_\alpha\hat{L}_\alpha)} \label{eq:redelta}
\end{equation}
and
\begin{equation}
\hat{U}(g(t))
={\rm e}^{-\ii t (\sum_{\alpha} (g_o)_\alpha\hat{L}_\alpha)}
  \label{eq:group}
\end{equation}
is the operator representation of the Lie group.

Thus, the tomogram can be rewritten in a form that only depends on group element
parameters $(g_0)_\alpha$ and the Radon variable $X$:
\begin{eqnarray}
\CW(X, (g_0)_\alpha) &=& \left\langle \delta\left(X\hat{\mathbb{I}}
- \sum_{\alpha} (g_o)_\alpha\hat{L}_\alpha\right)\right \rangle\cr
&=&\frac1{2\pi}{\rm Tr}\int \d t \, \hat{\rho}\,{\e}^{\ii tX}\,
\hat{U}(g(t)) \cr &=&\frac1{2\pi}\int \d t \, {\e}^{\ii tX}\, {\rm
Tr} \left(\hat{\rho}\, \hat{U}(g(t))\right).  \label{eq:groupfinale}
\end{eqnarray}
Therefore, the tomogram is related to the orbit of the group and the
Fourier integral of the trace of the orbit in the group
representation. This group-theoretical representation permits the
extension to more general cases \cite{Ibort}.

\section{Thick quantum tomography}
\label{sec:thick}

We now turn our attention to {\it thick} tomographic maps
\cite{tomothick}, which is a more realistic approach for practical
applications, because instead of marginals defined over lines, as in
the classical Radon transform \cite{Rad1917} or quadrics as in the 
quadratic generalized Radon transform  \cite{tomogram,tomocurved}, it involves a {\it thick}
window function $\Xi$. This is convoluted with the tomographic map
and concentrates the marginals around some given background curves
(that can be lines or quadrics), without resorting to a singular
delta function. For example, if the weight function $\Xi$ is a step
function, it defines marginals along thick lines or thick quadratic
curves. In the quantum case, this amounts to replacing in the
definition of the dequantizer $\hat{U}(x)$ the Dirac delta-function
by the weight function $\Xi$,
\begin{equation}
\hat{U}(X,{(g^0)_\alpha})= \Xi\left(X\hat{\mathbb{I}} -
\sum_{\alpha} (g_o)_\alpha\hat{L}_\alpha\right).
\label{dequantizertris}
\end{equation}
For the symplectic quantum tomography, one has the dequantizer
\begin{equation}
\hat{U}(X,{\mu},{\nu})= \Xi\left(X\hat{\mathbb{I}} -\mu \hat{q}-\nu
\hat{p}\right). \label{symdequantizerbis}
\end{equation}
The new tomogram reads
\begin{eqnarray}
\CW_{\Xi}(X,\mu,\nu) = {\rm Tr}\, \hat{\rho}\
\Xi\left(X\hat{\mathbb{I}} -\mu \hat{q}-\nu \hat{p}\right),
\label{eq:radonthick}
\end{eqnarray}
Using the Weyl map one obtains  a tomogram for the Wigner function
\begin{equation}
\CW_{\Xi}(X,\mu,\nu) = \frac1{2\pi}\int W(p,q) \,  \Xi\left(X -\mu {q}-\nu {p}\right)\; \d p\, \d q.
\label{radondos4}
\end{equation}
The interesting property of the above formula (\ref{radondos4}) is
that it can be inverted in complete analogy with the classical thick
tomography introduced in \cite{tomothick}. The thick tomogram can be
expressed in terms of standard symplectic tomograms via a
convolution formula
\begin{equation}
\CW_{{\Xi}}(X,\mu,\nu) = \int \CW(Y,\mu,\nu) \, \Xi\left(X -Y\right)\; \d Y,
\label{radondos5}
\end{equation}
which leads to the explicit construction of the inverse transform.
Indeed, the inverse transform is obtained by means of a Fourier transform of the convolution
integral
\begin{equation}
W({p},{q})= \frac{\mathcal{N}_{\Xi}}{2\pi} \int
\CW_{\Xi}(X,{\mu},{\nu}) \, {\rm e}^{\ii (X-\mu{q}-{\nu}{p})}\;
\d{X}\,\d{\mu}\, \d{\nu}, \label{iradonsim}
\end{equation}
where
\begin{equation}
\mathcal{N}_{\Xi}=\frac{1}{\widetilde{\Xi}(-1)}, \qquad  \widetilde{\Xi}(-1)=
 \int {\Xi}(z)  \, {\rm e}^{\ii z}\, \d{z}.\nonumber
\end{equation}
In invariant form, the state reconstruction is achieved by
\begin{eqnarray}
\hat{\rho}= \frac{\mathcal{N}_{\Xi}}{2\pi}
\int \CW_{\Xi}(X,{\mu},{\nu}) \,
{\rm e}^{\ii (X\mathbb{I}-\mu\hat{q}-{\nu}\hat{p})}\; \d{X}\,\d{\mu}\, \d{\nu}.\nonumber
\label{iradonsimm}
\end{eqnarray}
The quantizer operator in thick symplectic tomography is
\begin{eqnarray}
\hat{D}(X,{\mu},{\nu})=
\frac{\mathcal{N}_{\Xi}}{2\pi} \,
{\rm e}^{\ii (X\mathbb{I}-\mu\hat{q}-{\nu}\hat{p})}.
\label{qsim}
\end{eqnarray}
Let us now consider a particular example of thick tomogram to illustrate
the potentialities of the new method.  If the weight function is a gaussian function
\begin{equation}
\Xi(z)=\frac1{\sqrt{2\pi\sigma^2}}
{\rm e}^{-\frac {z^2}{2\sigma^2}}\nonumber
\end{equation}
which tends to the  delta distribution in the  $\sigma\to0$ limit,
\begin{equation}
\lim_{\sigma\to 0}\Xi(z)= {\delta}(z), \nonumber
\end{equation}
the thick tomogram of the coherent  states $|\alpha\rangle\langle\alpha|$ read
\begin{equation}
\CW_{\sigma}^\alpha(X,\mu,\nu) = \frac1{\sqrt{\pi(\mu^2+\nu^2+2\sigma^2)}}
{\rm e}^{-\frac{(X-\bar{X})^2}{\mu^2+\nu^2+2\sigma^2}},
\label{gauss}
\end{equation}
where
\begin{equation}
\bar{X}=\sqrt{2}\, \mu\, {\rm Re}\, \alpha+\sqrt{2}\,\nu\, {\rm Im}\, \alpha.
\end{equation}
For the vacuum state $|0\rangle\langle0|$, the tomogram reads
\begin{equation}
\CW_{\sigma}^{\rm vac}(X,\mu,\nu) =
\frac1{\sqrt{\pi(\mu^2+\nu^2+2\sigma^2)}} {\rm
e}^{-\frac{X^2}{\mu^2+\nu^2+2\sigma^2}}. \label{gaussvac}
\end{equation}
The quantizer reads
\begin{eqnarray}
\hat{D}_\sigma(X,{\mu},{\nu})= \frac1{2\pi}{\rm e}^{\frac{\sigma^2}{2}+\ii (X\mathbb{I}-\mu\hat{q}-{\nu}\hat{p})} 
\label{qqsim}
\end{eqnarray}
and the dequantizer is given by
\begin{eqnarray}
\hat{U}_\sigma(X,{\mu},{\nu})= \frac1{\sqrt{2\pi\sigma^2}}{\rm e}^{-\frac{(X\mathbb{I}-\mu\hat{q}-{\nu}\hat{p})^2}{2\sigma^2}}.
\label{dqsim}
\end{eqnarray}
One interesting property, that is preserved by the smoothing of the tomogram,
is that the marginals $\CW_{\Xi}(X,\mu,\nu)$ are also probability distributions.
In the  limit $\sigma\to 0$, $\Xi(z)\to \delta(z)$, $\widetilde{\Xi}(-1)=
1$, $\mathcal{N}_{\Xi}=1$.

In the case of multimode systems with quadratic Hamiltonians, the
dequantizer and quantizer of thick tomography with a Gaussian weight
function $\Xi$ become
\begin{eqnarray}
\hat{U}_\sigma(X,\vec{\mu},\vec{\nu})= \Xi\left(X\mathbb{I}-
\frac12(\hat{\vec{Q}}-\vec{r}) \cdot B (\hat{\vec{Q}}-\vec{r})\right) 
\label{dqqsim}
\end{eqnarray}
and
\begin{equation}
\hat{D}_\sigma(X,\vec{\mu},\vec{\nu}) = \frac{\left|\det {B}\right|}{(2\pi)^{N+1}
\tilde{\Xi}(-1)} {{\rm e}^{\ii \left(X\mathbb{I}-\frac12(\hat{\vec{Q}}-\vec{r}) \cdot B (\hat{\vec{Q}}-\vec{r})\right)},}
\label{qqqsim}
\end{equation}
respectively.
Thus, the quantum thick tomogram of a multimode state with  weight function
$\Xi$ reads
\begin{eqnarray}
\CW_\Xi(X,\vec{\mu},\vec{\nu})&=& \int\, \frac{d\vec{p}\, d\vec{q}}
{(2\pi)^N} W(\vec{q},\vec{p})\nonumber \\
&&  \Xi\left(X\mathbb{}-\frac12({\vec{Q}}-\vec{r}) B ({\vec{Q}}-\vec{r})\right),
\label{wwig}
\end{eqnarray}
in terms of the Wigner functions of the state, and the inverse formula
\begin{eqnarray}
W(\vec{q},\vec{p})&=& \frac{\left|\det {B}\right|}{(2\pi)^{N+1} \widetilde{\Xi}(-1)}\int\, {d X\, d\vec{\mu}\, d\vec{\nu}}
\,  \CW_\Xi(X,\vec{\mu},\vec{\nu})\nonumber \\
&&  {\rm e}^{i\left(X-\frac12({\vec{Q}}-\vec{r}) B ({\vec{Q}}-\vec{r})\right)}
\label{wwwig}
\end{eqnarray}
permits to reconstruct the Wigner function from its tomograms.

The proof of the inverse formula follows the same steps of
the case of linear symplectic tomograms. Indeed, if we plug  (\ref{wwig})
into the right hand side of (\ref{wwwig}) we get
\begin{eqnarray}
&&\!\!\!\!\! \!\!\!\!\! \frac{\left|\det {B}\right|}{(2\pi)^{N+1}
\widetilde{\Xi}(-1)}\int\, {d X\, d\vec{\mu}\, d\vec{\nu}}
\,  \int\, \frac{d\vec{p}\,'\, d\vec{q}\,'}
{(2\pi)^N} W(\vec{q}\,',\vec{p}\,')\nonumber \\
&&\!\!\!\!\!\!\!\!\!\!  \Xi\left(X\mathbb{}-\frac12({\vec{Q}'}
-\vec{r}) B ({\vec{Q}'}-\vec{r})\right)
  {\rm e}^{i\left(X-\frac12({\vec{Q}}-\vec{r}) B ({\vec{Q}}-\vec{r})\right)}\nonumber \\
&&\!\!\!\!\! \!\!\!\!\! =\frac{\left|\det {B}\right|}{(2\pi)^{N+1}
\widetilde{\Xi}(-1)}\int\, {d X\, d Y\, d\vec{\mu}\, d\vec{\nu}}
\,  \int\, \frac{d\vec{p}\,'\, d\vec{q}\,'} {(2\pi)^N} W(\vec{q}\,',\vec{p}\,')\nonumber \\
&& \delta(Y-\frac12({\vec{Q}'}-\vec{r}) B ({\vec{Q}'}-\vec{r}))
  \Xi\left((X-Y)\mathbb{I}\right)\nonumber \\
 && {\rm e}^{i\left(Y-\frac12({\vec{Q}}-\vec{r}) B ({\vec{Q}}-\vec{r})\right)}\ {\rm e}^{i(X-Y)}\nonumber \\
 &&\!\!\!\!\! \!\!\!\!\! =\frac{\left|\det {B}\right|}{2\pi^{}  \widetilde{\Xi}(-1)}\int\,
  {d Z\, d\vec{p}\,'\, d\vec{q}\,'}\, W(\vec{q}\,',\vec{p}\,')
\,  \int\, \frac{d\vec{\mu}\, d\vec{\nu}} {(2\pi)^N} \, \Xi\left(Z\mathbb{}\right)\, {\rm e}^{iZ}\nonumber \\
 && {\rm e}^{i\left(\vec{r} B ({\vec{Q}-\vec{Q}'})-\frac12\vec{Q}B\vec{Q}+\frac12\vec{Q}'B\vec{Q}'\right)}\nonumber \\
 &&\!\!\!\!\! \!\!\!\!\! =\frac{1}{\widetilde{\Xi}(-1)}\,   W(\vec{q},\vec{p})\int\, \frac{d Z}{2\pi}\, \Xi(Z) \, {\rm e}^{iZ}=
 W(\vec{q},\vec{p}) 
\label{wwwwig}
\end{eqnarray}

Let us consider the example of a quadratic thick tomogram governed
by a Hamiltonian with
\begin{equation}
B=\pmatrix{{\mathbb{I}_2},{\,0}\crcr
{\,\,\,  0\,\, },{\mathbb{I}_2}}
\label{nondeg}
\end{equation}
and Gaussian weight function $\Xi_\sigma$ for a system with one
degree of freedom. In that case, one might think that the tomogram
is centered on a quadratic curve
 \begin{equation}
 X= \frac12({{p}}-\nu)^2+ \frac12({{q}}-\mu)^2.
 \end{equation}
The tomogram is given by
\begin{eqnarray}
\CW_\sigma(X,{\mu},{\nu})&=&{\frac1{2\pi{\sqrt{2\pi \sigma^2}}}} \int\, {\d{p}\, \d{q}}\;
 W({q},{p})\nonumber \\
&&  {{\rm e}^{-\frac{\left(X- \frac12({{p}}-\nu)^2- \frac12({{q}}-\mu)^2\right)^2} {2\sigma^2}}} ,
\label{rrwwwig}
\end{eqnarray}
in terms of the  Wigner function and the  inverse transform
\begin{eqnarray}
W({q},{p})&=&\frac{{\rm e}^{\frac{\sigma^2}{2}}}{2\pi} \int\, {\d X\, \d{\mu}\, \d{\nu}}\; \CW_\sigma(X,{\mu},{\nu})
 \nonumber \\
&&  {{\rm e}^{\ii \left(X- \frac12({{p}}-\nu)^2- \frac12({{q}}-\mu)^2\right)}}{},
\label{wwwwig1}
\end{eqnarray}
allows a reconstruction of the Wigner function from its thick quadratic
tomograms.

\section{Comparison with the thick Radon transform}
\label{sec-thickRadon}

A comparison with the quantum Radon transform (\ref{homodyne}) is
useful and clarifies the merits of the approach taken in this
article. Let us consider the thick version of (\ref{homodyne}), with
a window function $\Xi$
\begin{equation}
\CR_{\Xi}(X,\varphi) =\tr \hat{\rho}\ \Xi (X \hat{\mathbb{I}} -\hat{q} \cos\varphi - \hat{p} \sin \varphi ) .
 \label{thickhomodyne}
\end{equation}
Now, in general, relation~(\ref{eq:m2radon}) (a consequence of the
positive homogeneity of degree $-1$ of the Dirac distribution) does
not hold anymore for $\CR_{\Xi}$ and $\CW_{\Xi}$. As a consequence,
one has to deconvolve $\CR_{\Xi}$---whenever this is possible---in
order to get $\CR_{\hat\rho}$ and thus $\hat\rho$. Since
\begin{equation}
\CR_{\Xi}(X,\varphi)=\int \Xi(X-Y) \CR_{\hat{\rho}}(Y,\varphi)\, \d Y,
\end{equation}
one gets
\begin{equation}
\int \d X\, \CR_{\Xi}(X,\varphi)\, \e^{-\ii k X} = \tilde{\Xi}(k) \int \d X\, \CR_{\hat{\rho}}(X,\varphi)\, \e^{-\ii k X}.
\label{eq:conv}
\end{equation}
Equation~(\ref{eq:conv}) can be inverted for any $\CR_{\hat{\rho}}$ if and only if  $\tilde{\Xi}(k)\neq 0$ for all $k$.
In this case, from~(\ref{eq:conv}) and (\ref{im2sim}) one gets
\begin{eqnarray}
\hat\rho  &=&
\frac{1}{2\pi }
\int \d X \int_0^\infty   \d r \int_0^{2\pi} \d \varphi\; \frac{\CR_{\Xi} (X,{\varphi })}{\tilde{\Xi}(-r)}
\nonumber \\
&&\times r\, \e^{\ii r(X \hat{\mathbb{I}} -\hat{q} \cos\varphi - \hat{p} \sin \varphi )}.
 \label{im2sim1}
\end{eqnarray}
Therefore, in contrast with the thick symplectic transform, the
thick Radon transform cannot be inverted for arbitrary window
functions. Moreover, even when the inversion is possible, one needs
to have a complete knowledge of the window function in order to
deconvolve the thick homodyne tomogram and perform its inversion
(\ref{im2sim1}).

\section{Conclusions}
\label{sec-concl} There are many possible interesting applications
of the quantum tomograms introduced in this article by using
quadrics and window functions in the phase-space. 
The measurements of variables that are quadratic forms of
positions and momenta are equivalent to measuring the energy of
vibrations with changing parameters, like minima of potential
energy, elastic forces, and so on. This procedure can be used to
measure the state $\hat{\rho}$ of trapped ions by varying the
parameters of the trapping potential.

Homodyne detection, by using non-linear crystals as beam splitters,
also creates quadratic forms of the photon quadrature components
which, in principle, can be measured by varying the parameters of
the quadratic forms. Thus, tomography with the generalized quantized
version of the Radon transform provides new possibilities to measure
quantum states.

From a formal viewpoint, the new  tomographic maps and their inverse
can be formulated within the framework of the star-product scheme,
with quantizers and dequantizers for multidimensional systems. Some
of the constructions involve  generic quadratic forms of the
different position and momentum operators. The inhomogeneous
symplectic group of the corresponding multidimensional phase spaces
and its subgroups play a significant role in the formalism. The new
quantum tomograms can be considered as the quantum version of the
generalized classical tomograms introduced in our previous work
\cite{tomocurved} and involve quadratic surfaces on phase
spaces. The correspondence between both approaches is established by
means of the Weyl map. 

In quantum optics, the new {\it non-linear}  Radon transforms can be
easily extended to the quantum domain by using the Weyl-Wigner map.
The results of this article show that the reconstruction of the
Wigner function using optical or symplectic tomography based on
straight-line Radon transform can be extended to situations in which
the marginals in phase space are measured for curved hyperbolas or
ellipses. In particular, parabolic tomography could be implemented
with the recently observed accelerated Airy beams \cite{airy}.

Finally, the other generalization introduced in this paper, named ``thick" quantum tomography, entails that
the effect of a non-singular window function not only smooths the tomograms, but
in fact matches better all practical applications, where thick tomograms arise in a natural way. 
The striking fact we pointed out in this work is that the reconstruction procedure
in this case is almost independent on the window functions, with an
enormous advantage due to the fact for most experimental devices the window
function is not completely known.

\acknowledgments

V.I.M. was partially supported by the Russian Foundation for Basic
Research under Project No.~11-02-00456. This work partially
supported by a cooperation grant INFN-MICINN. M.A. was also partially
supported by the Spanish MICINN grant FPA2009-09638 and DGIID-DGA
(grants 2009-2010-2011-E24/2).


\end{document}